\documentclass[conference]{IEEEtran}
\IEEEoverridecommandlockouts
\usepackage{cite}
\usepackage{amsmath,amssymb,amsfonts}
\usepackage{algorithmic}
\usepackage{graphicx}
\usepackage{textcomp}
\usepackage{xcolor}

\usepackage{booktabs}
\usepackage{array}
\usepackage{tabularx}
\usepackage[caption=false,font=footnotesize]{subfig}

\usepackage{float}
\usepackage{url}
\usepackage[hidelinks]{hyperref}

\usepackage[section]{placeins}

\def\BibTeX{{\rm B\kern-.05em{\sc i\kern-.025em b}\kern-.08em
    T\kern-.1667em\lower.7ex\hbox{E}\kern-.125emX}}
\begin{document}

\title{Model-intrinsic, Weight-level, and Training-free Spectral Fingerprinting for Large Language Models}

\author{
\IEEEauthorblockN{Yanan He$^1$, Yanli Chen$^{1,\star}$, Ling Xiong$^2$, Gejian Zhao$^3$ and Hanzhou Wu$^{1,3,\star}$}
% \thanks{$^\star$\emph{Corresponding authors: Y. Chen (yanli\_027@163.com), H. Wu (h.wu.phd@ieee.org)}}}
\IEEEauthorblockA{
$^1$School of Big Data and Computer Science, Guizhou Normal University, Guiyang 550025, China\\
$^2$School of Information Science, Beijing Language and Culture University, Beijing 100083, China\\
$^3$School of Communication and Information Engineering,
Shanghai University, Shanghai 200444, China\\
$^\star$Corresponding authors (e-mails: yanli\_027@163.com, h.wu.phd@ieee.org)}}
% 2934069038@qq.com, yanli\_027@163.com, lingdonghua99@163.com, 23820171@shu.edu.cn, h.wu.phd@ieee.org

\maketitle

\begin{abstract}
The increasing release and reuse of Large Language Models (LLMs) as open-weight checkpoints make model fingerprinting important for ownership protection and offline auditing. Existing fingerprinting methods typically construct evidence from query responses, generated outputs, or input-induced activations. Although effective in representation-based settings, they can be affected by prompt selection, decoding configuration, inference environment, and forward-pass implementation, which may limit their applicability. Here we introduce a model-intrinsic weight-level framework that derives compact fingerprints from spectral structures encoded in LLM parameters. Specifically, the framework constructs GATE spectral scalar features from the singular-value distributions of GATE projection weights in gated feed-forward networks (FFNs), and aggregates the resulting layer-wise features through depth-wise segmented pooling to form a model-level fingerprint. Such representation preserves coarse layer-depth organization while reducing dependence on strict layer-to-layer alignment, enabling comparison across LLMs with different depths. The similarity between fingerprints is computed by linear centered kernel alignment. The entire pipeline is training-free and does not require input samples, generated outputs, forward activations, or an auxiliary verifier. Experiments on representative LLM families and post-training variants have demonstrated a clear separation between source-related models and unrelated ones, and stable similarity under instruction tuning, quantization, pruning, and distillation-related modifications, which verify that spectral features are robust evidence for LLM fingerprinting. 
\end{abstract}
\begin{IEEEkeywords}
Large language models, weight-level fingerprint, provenance verification, ownership protection, spectral analysis.
\end{IEEEkeywords}

\section{Introduction}
\label{sec:introduction}
Large language models (LLMs) nowadays have been increasingly released, adapted, and redistributed as open-weight checkpoints. As these models become valuable intellectual assets, there is a growing need for model fingerprinting techniques that support ownership protection, provenance analysis, and offline auditing. In such scenarios, when model parameters are available, an auditor may examine whether a suspect checkpoint is related to a specified source model, motivating the development of fingerprinting methods that extract reproducible evidence directly from model weights.

Mainstream copyright protection and ownership verification methods mainly include digital watermarking and fingerprinting. Model watermarking typically embeds information within a model or into its generated content for subsequent ownership verification. The watermark information can be incorporated into model parameters, network structures, internal representations, model behaviors, or generated outputs. In the context of LLMs, watermarking has also been applied to generated texts, model weights, and other model-side representations. In parallel, model
fingerprinting characterizes a model through intrinsic behavioral, representational, or parameter-level evidence for identification and provenance analysis.

LLM fingerprinting methods differ in their evidence and verification settings. Recent black-box methods exploit model-specific subspaces for API-protected LLMs \cite{hiddenfingerprint}, perturbation-response distributions \cite{beyondqa}, concept-response patterns \cite{cref}, and robust verification against composite attacks \cite{robustblackbox}. Other methods use instruction tuning \cite{xu2024instructional} or linguistic/output behavior \cite{jin2024proflingo}, while merge-resistant verification \cite{yamabe2025mergeprint} and active query-response identification \cite{pasquini2025llmmap} target practical reuse and deployment. Earlier research also explored decision-boundary fingerprints \cite{cao2021ipguard}, conferrable adversarial fingerprints \cite{lukas2021conferrable}, and testing-based copyright protection frameworks \cite{chen2022copy}. For LLMs, fingerprints
have also been derived from weight-level invariants
\cite{zeng2024huref} and internal feature representations
\cite{zhang2025reef}. These methods provide useful evidence under different access assumptions: query-based fingerprints support black-box verification, representation-based methods require model execution and internal activations, while open-weight checkpoints
generally enable weight-level fingerprinting. Compared with behavior-based approaches, weight-level fingerprinting does not require carefully crafted prompts or stochastic decoding, making it suitable for parameter-level provenance analysis of open-weight LLMs. However, it remains challenging to construct a
compact fingerprint that distinguishes unrelated models while preserving source-related information under common post-training modifications. This motivates us to further investigate weight-level
fingerprinting for open-weight LLMs.

In modern Transformer-based LLMs, gated feed-forward networks (FFNs) are an important component of Transformer blocks. A gated FFN typically contains two parallel projection branches operating on the same hidden representation. The UP projection maps the input
into a higher-dimensional intermediate space, while the GATE projection generates an input-dependent gating signal that selectively modulates the expanded features through element-wise multiplication. The gated activations are subsequently projected back to the hidden dimension through a DOWN projection. Since the GATE branch directly participates in the modulation of intermediate features, its weight matrix provides a representative parameter-level object for analyzing internal model structure.

Building upon this observation, we investigate whether spectral features constructed from the singular-value distributions of GATE projection weights can serve as a model-intrinsic basis for LLM fingerprinting. Specifically, we extract a scalar spectral feature
from the GATE projection of each layer and aggregate the resulting layer-wise features along model depth to form a compact model-level fingerprint. The entire fingerprint extraction process operates directly on static model weights and requires neither input samples, model outputs, nor forward inference. Experiments show that the resulting GATE spectral fingerprints maintain high similarity among source-related models while providing clear separation from unrelated models, offering model-intrinsic, weight-level, and training-free evidence for LLM provenance verification.

\section{Preliminaries}
\label{sec:pre}
\subsection{Large Language Models}
Large language models are deep neural networks pretrained on large-scale text corpora to model distributions over token sequences and support a broad range of language understanding and generation tasks. In terms of the sequence-modeling backbone, LLMs can be constructed using Transformer, recurrent, state-space, or hybrid architectures, where Transformer is the most popular. Among Transformer-based language models, three common architectural forms are encoder-only, encoder-decoder, and decoder-only. Encoder-only models employ bidirectional self-attention to learn the contextual representations, encoder-decoder models perform conditional sequence modeling through separate encoding/decoding components, whereas decoder-only models generate tokens auto-regressively under causal self-attention, where the causal mask prevents information flow from future positions during training and inference.

This work focuses on decoder-only models equipped with gated FFNs. The target source-model families considered for provenance verification adopt this architecture, in which Transformer blocks are repeatedly stacked along model depth. Each block contains a separate set of attention and FFN parameters, providing a consistent layer-wise organization of projection weight matrices for weight-level provenance analysis.

Technically, given a sequence of $T$ tokens, the token embedding layer maps the discrete tokens to hidden representations, which are subsequently processed by a stack of $L$ Transformer blocks~\cite{vaswani2017attention}. Each Transformer block consists of a causal self-attention sublayer and a so-called position-wise FFN, together with residual connections and normalization layers. The causal self-attention sublayer models contextual dependencies among the current and preceding tokens, and the position-wise FFN applies the identical nonlinear transformation independently at each token position.

For a specific token position, let $x^{(\ell)} \in \mathbb{R}^{1\times d_m}$ represent the hidden representation entering the FFN in the $\ell$-th Transformer block, where $\ell=1,2,\ldots,L$ and $d_m$ means the model hidden dimension. After the final Transformer block, the output projection maps the hidden representations to vocabulary logits. For the target source-model families considered in this paper, the position-wise FFN is parameterized as a gated FFN.

\subsection{Gated Feed-Forward Networks}
A gated FFN contains two parallel input projections, i.e., the UP and GATE projections, followed by a DOWN projection. Given the hidden representation
$x^{(\ell)}$
entering the gated FFN in the $\ell$-th Transformer block, the three projections are
\begin{equation}
\left\{
\begin{aligned}
u^{(\ell)} &= x^{(\ell)} W_{\mathrm{up}}^{(\ell)}\\
g^{(\ell)} &= \mathcal{A}\left(x^{(\ell)} W_{\mathrm{gate}}^{(\ell)}
\right)\\
y^{(\ell)} &= \left(u^{(\ell)} \odot g^{(\ell)}\right) W_{\mathrm{down}}^{(\ell)}
\end{aligned}
\right.
\label{eq:gated_ffn}
\end{equation}
where $\mathcal{A}(\cdot)$ denotes the nonlinear activation function applied to the GATE branch, and $\odot$ denotes element-wise multiplication. The projection matrices satisfy that $W_{\mathrm{up}}^{(\ell)}\in\mathbb{R}^{d_m\times d_f}$, $W_{\mathrm{gate}}^{(\ell)}\in\mathbb{R}^{d_m\times d_f}$, and $W_{\mathrm{down}}^{(\ell)}\in\mathbb{R}^{d_f\times d_m}$, where $d_m$ and $d_f$ denote the model hidden dimension and the intermediate FFN dimension, respectively. Therefore, both $u^{(\ell)}$ and $g^{(\ell)}$ have $d_f$ components, and $y^{(\ell)}$ has $d_m$ components. The UP projection maps the hidden representation to a higher-dimensional intermediate space to increase representational capacity, while the GATE projection generates a data-dependent gating signal that selectively modulates the expanded features through element-wise multiplication. The DOWN projection maps the gated intermediate representation back to the model hidden dimension. Since the GATE branch generates the modulation signal for the expanded intermediate features, its weight matrix provides a meaningful weight-level object for spectral analysis within each Transformer block.

\section{Model-intrinsic, Weight-level, and Training-free Spectral Fingerprinting Technique}
\label{sec:method}
\begin{figure*}[!t]
\centering
\includegraphics[width=\linewidth]{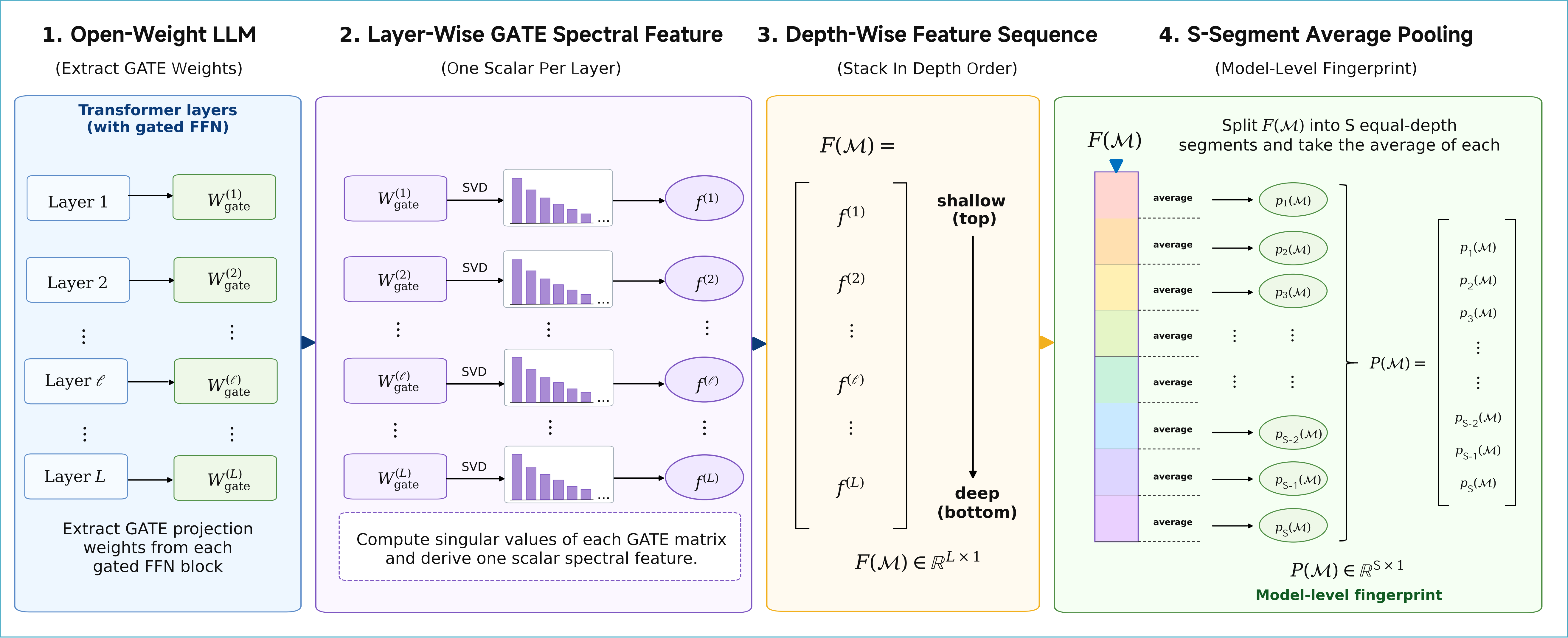}
\caption{General framework for the construction of the model-level GATE spectral fingerprint from layer-wise GATE projection weights.}
\label{fig:framework}
\end{figure*}

We construct the spectral fingerprint from layer-wise GATE spectral features. For each GATE projection, the distribution of singular values reflects how the weight matrix organizes its spectral energy across orthogonal transformation directions. We construct the GATE spectral feature from this distribution by firstly determining the spectral entropy of the normalized singular values and then exponentiating it. This feature summarizes whether the spectral energy of a GATE projection is concentrated in a few dominant directions or distributed over more directions. The resulting layer-wise features are further arranged along model depth and aggregated to produce a fixed-size model-level fingerprint for similarity analysis.

As shown in Fig.~\ref{fig:framework},  for each gated FFN block of a model $\mathcal{M}$, the GATE weight matrix is converted into a scalar $f^{(\ell)}$. Such layer-wise features are stacked in depth order as
\begin{equation*}
F(\mathcal{M})=[f^{(1)};f^{(2)};\ldots;f^{(L)}]\in\mathbb{R}^{L\times1}
\end{equation*}
and aggregated through segmented mean pooling to obtain
\begin{equation*}
P(\mathcal{M})=[p_1(\mathcal{M});p_2(\mathcal{M});\ldots;p_s(\mathcal{M})]\in\mathbb{R}^{s\times1},    
\end{equation*}
where $s$ is empirically set to 8 by default. Given two models, the similarity between them is computed by applying the linear centered kernel alignment (CKA) to their fingerprint vectors. 

\subsection{Layer-wise GATE Spectral Feature}
\label{subsec:layer_feature}
For the $\ell$-th gated FFN block, we extract a layer-wise spectral feature from the GATE weight matrix
$W_{\mathrm{gate}}^{(\ell)}\in\mathbb{R}^{d_m\times d_f}$. 
We perform singular value decomposition on this matrix and obtain its corresponding singular values, the number of which is
$q=\min(d_m,d_f)$.
To retain the dominant transformation modes and improve computational efficiency, we use the top-$r$ ($r\leq q$) singular values for subsequent feature construction.

The singular values characterize the transformation strength of the GATE projection along orthogonal directions, and their distribution reflects how the spectral energy of the matrix is organized. Let the retained ordered singular values of
$W_{\mathrm{gate}}^{(\ell)}$
be denoted as
$\{\sigma_{g,1}^{(\ell)},\sigma_{g,2}^{(\ell)},\ldots,\sigma_{g,r}^{(\ell)}\}$.
We normalize them as
\begin{equation}
\bar{\sigma}_{g,i}^{(\ell)}
=
\frac{
\sigma_{g,i}^{(\ell)}
}{
\sum_{j=1}^{r}\sigma_{g,j}^{(\ell)}+\epsilon
},
\ i=1,2,\ldots,r .
\label{eq:normalized_gate_spectrum}
\end{equation}
where $\epsilon$ is used for numerical stability. The normalized singular values form a distribution over the retained singular directions and describe the relative contribution of each direction.

Based on this normalized distribution, we first compute the spectral entropy as
\begin{equation}
H_g^{(\ell)}
=
-\sum_{i=1}^{r}
\bar{\sigma}_{g,i}^{(\ell)}
\log(\bar{\sigma}_{g,i}^{(\ell)}+\epsilon).
\label{eq:gate_entropy}
\end{equation}
The GATE spectral scalar feature is then defined by exponentiating this spectral entropy:
\begin{equation}
f^{(\ell)}
=
\exp(H_g^{(\ell)})\,\in\mathbb{R}.
\label{eq:layer_feature}
\end{equation}
Here, $H_g^{(\ell)}$ is an intermediate spectral-entropy statistic, while
$f^{(\ell)}$
is the final scalar feature used for fingerprint construction. This feature summarizes how the spectral energy of the GATE projection is concentrated or say dispersed across the retained singular directions. A smaller value indicates that the spectral energy is concentrated in fewer dominant directions, while a larger one indicates that the energy is distributed over more directions. Applying the identical procedure to all gated FFN blocks yields the layer-wise feature sequence $F(\mathcal{M})$.

\subsection{Model-level Fingerprint Construction}
The number of rows in $F(\mathcal{M})$ varies with model depth. Direct comparison of two layer-wise feature sequences therefore requires explicit layer correspondence between models with different numbers of gated FFN blocks. To obtain a fixed-size representation, the ordered block indices
$\{1,2,\ldots,L\}$ are partitioned into $s$ approximately equal-sized contiguous depth groups, denoted as $\{I_1,I_2,\ldots,I_s\}$, where $s | L$ generally. When $L$ is not divisible by $s$, the groups are constructed as approximately equal-sized contiguous depth segments. Each $I_k$ contains the layer indices belonging to the $k$-th relative-depth region, and $n_k=|I_k|$ denotes the number of layers in this group. The choice of 8 groups is used as the default setting and is examined in the ablation study. The features within the $k$-th group are mean-pooled as
\begin{equation}
p_k(\mathcal{M})
=
\frac{1}{n_k}
\sum_{\ell\in I_k}
f^{(\ell)},
\ k=1,2,\ldots,s.
\label{eq:mean_pooling}
\end{equation}
The $s$ features are stacked from shallow to deep to form
\[
P(\mathcal{M})
=
[p_1(\mathcal{M});p_2(\mathcal{M});\ldots;p_s(\mathcal{M})]
\in
\mathbb{R}^{s\times1}.
\]

This fingerprint construction removes the dependence on the model depth while preserving the coarse depth-wise organization of the GATE spectral-feature structure.

\subsection{Fingerprint Similarity}
Given two LLMs $\mathcal{M}_A$ and $\mathcal{M}_B$, there are different ways to determine the similarity between their model-level fingerprints, i.e., $P(\mathcal{M}_A)$ and $P(\mathcal{M}_B)$. Here we use linear CKA~\cite{kornblith2019similarity} for similarity comparison, which has been widely used for feature representation analysis of neural networks. 

The linear CKA between two \emph{centered} feature matrices $X\in \mathbb{R}^{n\times d_x}$ and $Y\in \mathbb{R}^{n\times d_y}$ is defined as
\begin{equation}
\text{CKA}(X,Y) = \frac{||X^\top Y||^2_\text{F}}{||X^\top X||_\text{F}||Y^\top Y||_\text{F}}.   
\end{equation}
Linear CKA determines the cosine similarity between the centered Gram matrices. Since $P(\mathcal{M}_A)$ and $P(\mathcal{M}_B)$ correspond to a column vector, it essentially requires us to find the square of the cosine similarity determined on mean-centered vectors, i.e., the squared Pearson correlation coefficient. This also implies that, in the one-dimensional feature setting, linear CKA measures the strength of the linear relationship between two variables while being insensitive to the sign of the correlation, treating positive and negative correlations equally.

\section{Experimental Results and Analysis}
\subsection{Setup}
We evaluate the proposed technique through pairwise model comparison
on publicly available LLM checkpoints. The experiments cover
representative open-weight LLM families, including
Qwen2.5\footnote{Online available:
\url{https://huggingface.co/Qwen/Qwen2.5-7B}},
Mistral\footnote{Online available:
\url{https://huggingface.co/mistralai/Mistral-7B-v0.3}},
Baichuan2\footnote{Online available:
\url{https://huggingface.co/baichuan-inc/Baichuan2-7B-Base}},
LLaMA\footnote{Online available:
\url{https://huggingface.co/meta-llama/Llama-3.1-8B}},
and several cross-family reference models. Table~\ref{tab:benchmarked_models}
summarizes the representative model families and checkpoints considered
in our experiments.

For pairwise evaluation, we consider three groups of model pairs.
The related group evaluates source-family consistency using
instruction-tuned, domain-specialized, chat, and other same-family
derivatives. The robust group evaluates fingerprint stability under
pruning, quantization, and distillation-related modifications, while
the unrelated group uses cross-family or cross-architecture pairs
to evaluate source separation.

\begin{table}[!t]
\centering
\caption{Representative open-weight LLM checkpoints considered in our experiments.}
\label{tab:benchmarked_models}
\scriptsize
\setlength{\tabcolsep}{2.8pt}
\renewcommand{\arraystretch}{1.10}
\begin{tabularx}{\columnwidth}{@{}l l l X@{}}
\toprule
\textbf{Group} & \textbf{Family} & \textbf{Source Model}
& \textbf{Compared Models} \\
\midrule
Related
& Qwen2.5
& Qwen2.5-7B
& Instruct, Coder, Math \\

Related
& Mistral
& Mistral-7B-v0.3
& Instruct-v0.3 \\

Related
& Baichuan2
& Baichuan2-7B-Base
& Chat \\

Related
& LLaMA
& LLaMA-3.1-8B
& Hermes-3, Aloe-Beta \\

\midrule
Robust
& Qwen2.5
& Qwen2.5-7B / Instruct
& Pruned-50p, GPTQ-Int4/Int8, Distill-Qwen \\

Robust
& Mistral
& Mistral-7B-v0.3
& Prune6 \\

\midrule
Unrelated
& Cross-family
& Qwen2.5 / Mistral / LLaMA
& Baichuan2, InternLM2.5, Chinese-LLaMA, Yi, ChatGLM2,
DeepSeek-Coder, GPT2-Large$^\dagger$ \\
\bottomrule
\end{tabularx}
\vspace{1pt}

{\scriptsize $^\dagger$ Auxiliary non-gated spectral analysis using
the FFN \texttt{c\_proj} projection.}
\end{table}

The 29-pair evaluation set contains 9 related pairs, 6 robust pairs,
and 14 unrelated pairs, and is used for detailed pairwise analysis,
robustness evaluation, and cross-method comparison. To further
evaluate the proposed fingerprint under broader model-pair diversity,
we construct an expanded set of 120 pairs over 25 checkpoints,
including 30 related and 90 unrelated pairs. The expanded set is used
for ROC and threshold analysis, truncation-rank sensitivity, and
structure-matched evaluation. For the 29-pair evaluation, we use
$r=128$ and $s=8$ for fingerprint construction. Since the truncation
rank $r$ controls the spectral information retained in the descriptor
and can affect the resulting verification performance, we do not
regard $r=128$ as a universally optimal setting; its influence is
further examined in the ablation study.

For the 29-pair evaluation, we report Related Mean, Related Min,
Robust Mean, Robust Min, Unrelated Mean, and Unrelated Max based
on the linear CKA similarities. The conservative separation gap is
defined as
\begin{equation}
\Delta
=
\mathrm{Related\ Min}
-
\mathrm{Unrelated\ Max}.
\end{equation}
For the expanded evaluation, we report AUROC, normalized pAUC@5\%
FPR, TPR@1\% FPR, and the Kolmogorov--Smirnov (KS) statistic.

\subsection{Main Results}
This subsection reports pairwise fingerprint comparisons on the
29-pair evaluation set. Table~\ref{tab:related_robust_results}
reports source-related and post-training modified pairs, whereas
Table~\ref{tab:unrelated_results} reports unrelated pairs from
different model families or architectures.

\begin{table}[!htbp]
\centering
\caption{Linear CKA similarities for source-related and post-training modified LLM pairs.}
\label{tab:related_robust_results}
\scriptsize
\setlength{\tabcolsep}{3.2pt}
\renewcommand{\arraystretch}{1.08}
\resizebox{\columnwidth}{!}{%
\begin{tabular}{@{}lllc@{}}
\toprule
\textbf{Group} & \textbf{Source Model} & \textbf{Compared Model}
& \textbf{Similarity} \\
\midrule
Related & Qwen2.5-7B & Qwen2.5-7B-Instruct & 1.0000 \\
Related & Qwen2.5-7B & Qwen2.5-Coder-7B & 0.9727 \\
Related & Qwen2.5-7B & Qwen2.5-Math-7B & 0.9862 \\
Related & Qwen2.5-Math-7B & Qwen2.5-Math-7B-Instruct & 0.9959 \\
Related & Qwen2.5-Coder-7B & Qwen2.5-Coder-7B-Instruct & 0.9999 \\
Related & Mistral-7B-v0.3 & Mistral-7B-Instruct-v0.3 & 1.0000 \\
Related & Baichuan2-7B-Base & Baichuan2-7B-Chat & 0.9997 \\
Related & LLaMA-3.1-8B & Hermes-3-Llama-3.1-8B & 1.0000 \\
Related & LLaMA-3.1-8B & Llama3.1-Aloe-Beta-8B & 0.9999 \\
\midrule
Robust & Qwen2.5-7B & Qwen2.5-7B-pruned-50p & 1.0000 \\
Robust & Qwen2.5-7B-Instruct & GPTQ-Int4 & 0.9256 \\
Robust & Qwen2.5-7B-Instruct & GPTQ-Int8 & 0.9108 \\
Robust & Qwen2.5-7B-Instruct & DeepSeek-R1-Distill-Qwen-7B & 0.9846 \\
Robust & Mistral-7B-v0.3 & Mistral-7B-v0.3-prune6 & 0.9198 \\
Robust & Hermes-3-Llama-3.1-8B & Llama3.1-Aloe-Beta-8B & 0.9999 \\
\bottomrule
\end{tabular}%
}
\end{table}

As shown in Table~\ref{tab:related_robust_results}, source-related
models consistently obtain high similarities. The lowest related
similarity is $0.9727$, observed between Qwen2.5-7B and
Qwen2.5-Coder-7B, while most other related pairs are close to $1.0$.
This indicates that the GATE spectral fingerprint is well preserved
across the evaluated instruction-tuned, domain-specialized, chat,
and other same-family adapted variants.

Under the evaluated pruning, GPTQ quantization, and
distillation-related modifications, all robust-pair similarities are at least $0.9108$. In particular, Mistral-7B-v0.3-prune6
reduces the model depth from 32 to 26 layers while retaining a
similarity of $0.9198$, indicating robustness to the tested
structural pruning.

\begin{table}[!htbp]
\centering
\caption{Linear CKA similarities for unrelated LLM pairs from different families or architectures.}
\label{tab:unrelated_results}
\scriptsize
\setlength{\tabcolsep}{2.8pt}
\renewcommand{\arraystretch}{1.08}
\begin{tabularx}{\columnwidth}{@{}XXc@{}}
\toprule
\textbf{Source Model} & \textbf{Unrelated Model} & \textbf{Similarity} \\
\midrule
Qwen2.5-7B & Baichuan2-7B-Base & 0.2288 \\
Qwen2.5-7B & Mistral-7B-v0.3 & 0.1876 \\
Mistral-7B-v0.3 & Baichuan2-7B-Base & 0.0965 \\
Qwen2.5-7B & Llama3.1-Aloe-Beta-8B & 0.0132 \\
Qwen2.5-7B & Hermes-3-Llama-3.1-8B & 0.0152 \\
Qwen2.5-7B & InternLM2.5-7B & 0.0173 \\
Qwen2.5-7B & Chinese-LLaMA-2-7B-hf & 0.1422 \\
Mistral-7B-v0.3 & Yi-1.5-9B & 0.1775 \\
Mistral-7B-v0.3 & ChatGLM2-6B & 0.1421 \\
Mistral-7B-v0.3 & DeepSeek-Coder-6.7B-Base & 0.0318 \\
Mistral-7B-v0.3 & GPT2-Large$^\dagger$ & 0.0231 \\
Mistral-7B-v0.3 & Chinese-LLaMA-2-7B-hf & 0.2010 \\
LLaMA-3.1-8B & Qwen2.5-7B & 0.0157 \\
LLaMA-3.1-8B & Yi-1.5-9B & 0.0003 \\
\bottomrule
\end{tabularx}
\end{table}

As shown in Table~\ref{tab:unrelated_results}, unrelated pairs from
different model families or architectures remain low, with a maximum
similarity of $0.2288$, substantially below the minimum related
similarity of $0.9727$. This indicates clear separation between the
two groups. Structural effects are examined further using
structure-matched pairs in the expanded evaluation.

\begin{table}[!htbp]
\centering
\caption{Group-level summary of linear CKA similarities.}
\label{tab:summary_results}
\scriptsize
\setlength{\tabcolsep}{3.2pt}
\renewcommand{\arraystretch}{1.12}
\begin{tabular*}{\columnwidth}{@{\extracolsep{\fill}}lcccc@{}}
\toprule
\textbf{Pair Group} & \textbf{No.} & \textbf{Mean}
& \textbf{Min} & \textbf{Max} \\
\midrule
Related   & 9  & 0.9949 & 0.9727 & 1.0000 \\
Robust    & 6  & 0.9568 & 0.9108 & 1.0000 \\
Unrelated & 14 & 0.0923 & 0.0003 & 0.2288 \\
\bottomrule
\end{tabular*}
\end{table}

Table~\ref{tab:summary_results} shows mean similarities of $0.9949$
and $0.0923$ for the Related and Unrelated groups, respectively.
According to Eq.~(7), the corresponding conservative separation gap
is $0.7439$. The Robust group achieves a mean similarity of $0.9568$
and a minimum of $0.9108$.

\subsection{Ablation Study}
\begin{figure}[!htbp]
\centering
\includegraphics[width=0.95\columnwidth]{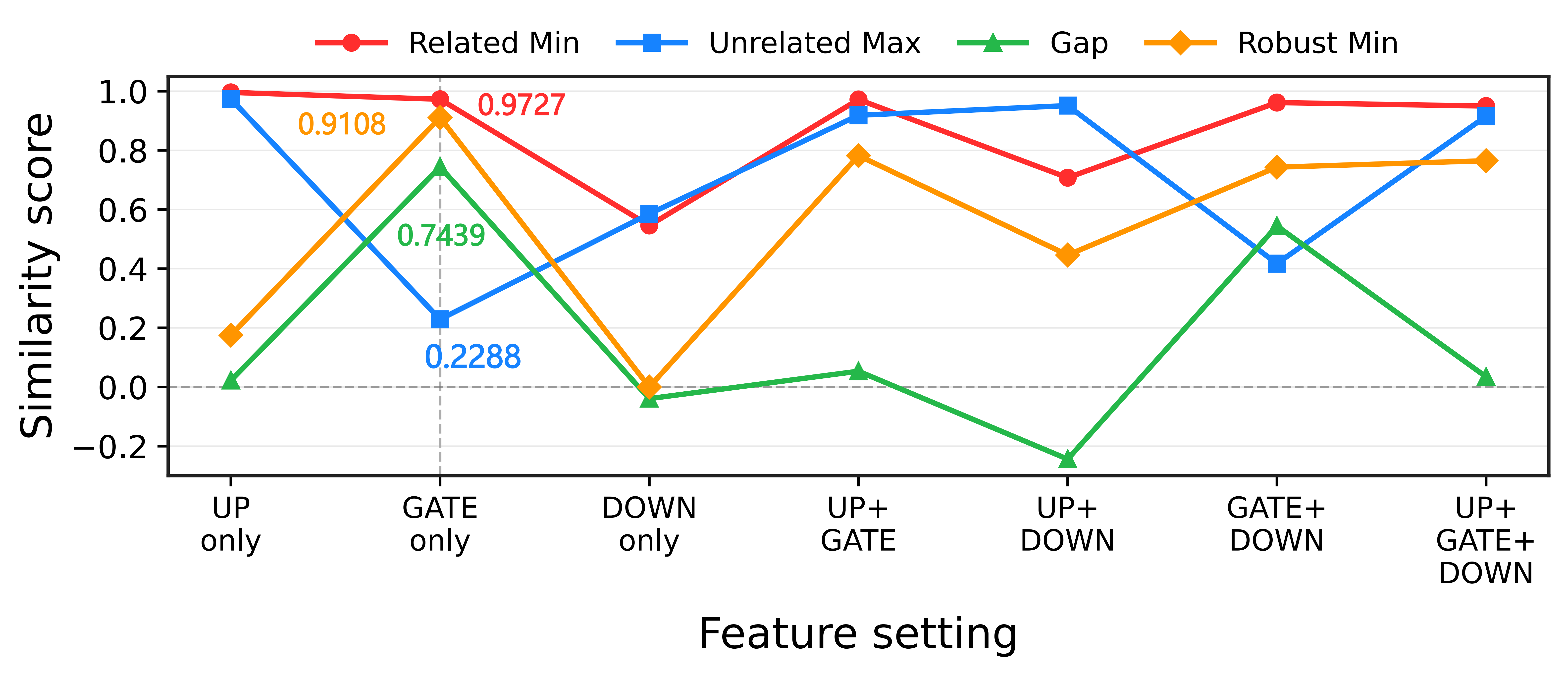}
\caption{Ablation of FFN projection sources on the 29-pair set
with $r=128$.}
\label{fig:feature_ablation}
\end{figure}

We analyze three design factors: the FFN projection source, the
number of depth segments, and the truncation rank. Projection-source and segment-number ablations use the 29-pair set with $r=128$, whereas rank sensitivity is evaluated on the expanded 120-pair set. Fig.~\ref{fig:feature_ablation} compares UP-only, GATE-only, DOWN-only, and their combinations. GATE-only provides a favorable balance between source separation and robustness, with an unrelated maximum of $0.2288$, a separation gap of $0.7439$, and a robust minimum of $0.9108$. UP-only yields higher unrelated similarities, while DOWN-only weakens related-model consistency; combining additional projections does not provide consistent improvements. We therefore adopt GATE as the fingerprint source.

The effect of the segment number is reported in
Table~\ref{tab:ablation_results}. Increasing $s$ generally reduces
unrelated similarity but also weakens robustness, with a substantial
drop at $s=16$. The setting $s=8$ provides a suitable balance in
the current experiments and is therefore used as the default.

\begin{table}[!htbp]
\centering
\caption{Segment-number ablation on the 29-pair set with $r=128$.}
\label{tab:ablation_results}
\scriptsize
\setlength{\tabcolsep}{3.2pt}
\renewcommand{\arraystretch}{1.12}
\begin{tabular*}{\columnwidth}{@{\extracolsep{\fill}}lcccc@{}}
\toprule
$s$ & \textbf{Rel. Min} & \textbf{Unrel. Max}
& \textbf{Gap} & \textbf{Rob. Min} \\
\midrule
4  & 0.9868 & 0.2697 & 0.7171 & 0.9169 \\
7  & 0.9791 & 0.2776 & 0.7015 & 0.8636 \\
8  & 0.9727 & 0.2288 & 0.7439 & 0.9108 \\
12 & 0.9683 & 0.2094 & 0.7589 & 0.6816 \\
16 & 0.9638 & 0.1802 & 0.7836 & 0.0318 \\
\bottomrule
\end{tabular*}
\end{table}

The truncation rank controls the number of leading singular values
retained for spectral-entropy computation. As shown in
Table~\ref{tab:rank_ablation}, performance varies non-monotonically
with $r$. The setting $r=64$ achieves the highest AUROC, pAUC@5\%,
and TPR@1\%, whereas $r=32$ gives the highest KS.

\begin{table}[!htbp]
\centering
\caption{Rank sensitivity on the expanded 120-pair set.}
\label{tab:rank_ablation}
\scriptsize
\setlength{\tabcolsep}{3.2pt}
\renewcommand{\arraystretch}{1.12}
\begin{tabular*}{\columnwidth}{@{\extracolsep{\fill}}lcccc@{}}
\toprule
$r$ & \textbf{AUROC} & \textbf{pAUC@5\%}
& \textbf{TPR@1\%} & \textbf{KS} \\
\midrule
32  & 0.9830 & 0.7407 & 0.5000 & 0.9111 \\
64  & 0.9870 & 0.8407 & 0.8333 & 0.9000 \\
128 & 0.9781 & 0.6926 & 0.4333 & 0.9000 \\
256 & 0.9448 & 0.7667 & 0.7667 & 0.7667 \\
\bottomrule
\end{tabular*}
\end{table}

On the original 29-pair set, both $r=64$ and $r=128$ achieve
$1.0000$ on all four discrimination metrics, indicating that the
preferred rank depends on the evaluation population. We therefore
retain $r=128$ for the detailed pairwise analysis as a fixed
experimental setting rather than a universally optimal choice.

\subsection{Expanded Verification and Robustness Analysis}
We further evaluate the proposed fingerprint on an expanded set of
120 model pairs, including 30 related and 90 unrelated pairs. At
$r=64$, the AUROC reaches $0.9870$. With a similarity threshold of
$0.9$, the TPR and FPR are $90.00\%$ and $4.44\%$, respectively;
the threshold is used only as an empirical operating point.

On a structure-matched subset with the same GATE-matrix dimensions
and network depth, the AUROC reaches $1.0000$ at $r=128$ and
$0.9800$ at $r=64$, indicating that these structural factors alone
do not determine fingerprint similarity. At $r=128$, we also inspect signed correlations over the 120 pairs: 23 pairs are negatively correlated, but none has a correlation at or below $-0.9$, and no abnormally high similarity caused by strong anticorrelation is observed.

We further evaluate targeted white-box spectral perturbations on
GATE weights using Qwen2.5-7B as the source and Qwen2.5-7B-Instruct as the target, with random seed 2026, under $r=128$ and $s=8$. Rank-1 and Rank-4 perturbations target one and four dominant singular directions, respectively, while $\epsilon$ denotes the relative Frobenius-norm perturbation budget. Budget-matched Gaussian perturbations are used
as a random control. Results are reported in Table~\ref{tab:attack_results}. As shown in
Table~\ref{tab:attack_results}, targeted perturbations affect the
fingerprint more strongly than random perturbations. At $\epsilon=2\%$, Rank-4 CKA drops to $0.6519$ with a $12.41\%$
increase in held-out perplexity, while the Gaussian control remains
near $1.0$. Under utility-constrained selection at the
$\epsilon=3\%$ candidate setting, the CKA is $0.8910$ with only a
$3.08\%$ increase in held-out perplexity, showing that stronger
fingerprint degradation generally incurs a utility cost.

\begin{table}[!htbp]
\centering
\caption{Robustness under targeted GATE spectral perturbations.}
\label{tab:attack_results}
\scriptsize
\setlength{\tabcolsep}{2.2pt}
\renewcommand{\arraystretch}{1.08}
\begin{tabular*}{\columnwidth}{@{\extracolsep{\fill}}ccccc@{}}
\toprule
$\epsilon$ &
\textbf{Rank-1 CKA} &
\textbf{Rank-4 CKA} &
\textbf{Gaussian CKA} &
\textbf{Rank-4 PPL Ratio} \\
\midrule
0.5\% & 0.9958 & 0.9798 & 1.0000 & 1.0021 \\
1.0\% & 0.9810 & 0.9206 & 1.0000 & 1.0169 \\
2.0\% & 0.9293 & 0.6519 & 1.0000 & 1.1241 \\
\bottomrule
\end{tabular*}
\end{table}

\subsection{Comparative Evaluation and Limitations}
We further compare the proposed method with representative LLM
fingerprinting and provenance-verification approaches.
Table~\ref{tab:method_comparison} summarizes their evidence sources
and access requirements. Compared with query- or output-based methods,
the proposed method requires neither prompt construction nor response
collection. Compared with representation-level methods, it also avoids
forward passes and input-dependent activations by extracting
fingerprints directly from static GATE weights.

\begin{table}[!htbp]
\centering
\caption{Methodological comparison with representative LLM fingerprinting methods.}
\label{tab:method_comparison}
\scriptsize
\setlength{\tabcolsep}{3pt}
\renewcommand{\arraystretch}{1.08}
\begin{tabularx}{\columnwidth}{@{}lXcc@{}}
\toprule
\textbf{Method} &
\textbf{Evidence Source} &
\textbf{Input/Fwd.} &
\textbf{Weight-only} \\
\midrule
Instructional FP &
Instruction--response behavior &
Yes & No \\

ProFLingo &
Black-box linguistic/output behavior &
Yes & No \\

MergePrint &
Behavioral fingerprint &
Yes & No \\

LLMmap &
Active query--response information &
Yes & No \\

HuRef &
Parameter invariants/reference evidence &
No & Yes \\

REEF &
Input-induced internal representations &
Yes & No \\

Ours &
GATE weight spectra &
No & Yes \\
\bottomrule
\end{tabularx}
\end{table}

For numerical comparison, we further evaluate REEF and adapted
HuRef invariant-terms cosine similarity (HuRef-ICS) on the same
29 model pairs, as reported in Table~\ref{tab:numerical_comparison}. As shown in
Table~\ref{tab:numerical_comparison}, our method and adapted
HuRef-ICS both achieve $1.0000$ on AUROC, pAUC@5\%, TPR@1\%,
and KS. Our method further yields a Mahalanobis separation of
$14.402$, indicating a stronger standardized separation between
related and unrelated score distributions in the current evaluation.

\begin{table}[!htbp]
\centering
\caption{Numerical comparison on the same 29 model pairs.}
\label{tab:numerical_comparison}
\scriptsize
\setlength{\tabcolsep}{2.8pt}
\renewcommand{\arraystretch}{1.08}
\begin{tabular*}{\columnwidth}{@{\extracolsep{\fill}}lccccc@{}}
\toprule
\textbf{Method} &
\textbf{AUROC} &
\textbf{pAUC@5\%} &
\textbf{TPR@1\%} &
\textbf{KS} &
\textbf{Mah.} \\
\midrule
Ours              & 1.0000 & 1.0000 & 1.0000 & 1.0000 & 14.402 \\
REEF              & 0.8857 & 0.8000 & 0.8000 & 0.8000 & 1.297 \\
Adapted HuRef-ICS & 1.0000 & 1.0000 & 1.0000 & 1.0000 & 2.859 \\
\bottomrule
\end{tabular*}
\end{table}

The standard fingerprint extraction relies on gated FFNs, and its
applicability to non-gated architectures has not yet been
comprehensively evaluated. Larger models, adversarial fine-tuning,
model merging, and fingerprint collision risks in large model
registries also require further study.

\section{Conclusion}
\label{sec:conclusion}
This paper presents a model-intrinsic, weight-level, and training-free spectral fingerprinting method for LLM provenance verification. It extracts GATE spectral features directly from static weights and compares compact depth-wise fingerprints using linear CKA.
Experiments demonstrate strong source separation and robustness under the evaluated model modifications, while expanded verification and baseline comparisons further support its effectiveness. Requiring neither input samples nor forward inference, the method provides a practical approach for offline LLM provenance analysis.

\section*{Acknowledgment}
\label{sec:ack}
This research was supported by the Basic Research Program for Natural Science of Guizhou Province under Grant Number QIANKEHEJICHU-ZD[2025]-043, the Science and Technology Commission of Shanghai Municipality under Grant Number 24ZR1424000, the Nanning ``Yong Jiang'' Program under Grant Number RC20250102, and the National Natural Science Foundation of China under Grant Number U23B2023.

\end{document}